\documentclass[aps,prd,onecolumn,groupedaddress,showpacs,showkeys,nofootinbib,amssymb]{revtex4}
\usepackage[T1]{fontenc}
\usepackage[latin1]{inputenc}
\usepackage{graphicx}
\usepackage[english]{babel}
\usepackage{amsmath}
\usepackage{amssymb}
\usepackage{amsfonts}
\usepackage{psfrag}

\def\be{\begin{equation}}
\def\ee{\end{equation}}
\def\bea{\begin{eqnarray}}
\def\eea{\end{eqnarray}}
\def\beq{\begin{eqnarray}}
\def\eeq{\end{eqnarray}}

\newtheorem{Proposition}{Proposition}[section]

\newfont{\gotico}{eufm10 scaled\magstephalf}
\newfont{\qvd}{msam10 scaled\magstephalf}

\def\de#1/de#2{\frac{\partial {#1}}{\partial {#2}}}
\def\De#1/de#2{\dfrac{\partial {#1}}{\partial {#2}}}

\def\const{{\rm const.}}

\makeindex \makeatletter

\def\widebar{\accentset{{\cc@style\underline{\mskip10mu}}}}
\makeatother

\begin{document}

\numberwithin{equation}{section}

\title{ The Cauchy problem for $f(R)$-gravity: an overview}

\author{S. Capozziello$^{1,2}$ and S. Vignolo$^{3}$ }

\affiliation{$~^{1}$ Dipartimento di Scienze Fisiche,
Universit\`{a} ``Federico II'' di Napoli and $~^{2}$INFN Sez. di Napoli,
Compl. Univ. Monte S. Angelo Ed. N, via Cinthia, I- 80126 Napoli
(Italy)}

\affiliation{$^{2}$DIPTEM Sez. Metodi e Modelli Matematici,
Universit\`a di Genova,  Piazzale Kennedy, Pad. D - 16129 Genova (Italy)}

\date{\today}

\begin{abstract}
 We review the Cauchy problem for $f(R)$ theories of gravity, in metric and metric-affine
 formulations, pointing out analogies and differences with respect to  General Relativity.  
The role of conformal transformations,  effective scalar fields and sources in the field equations is discussed in view of the well-posedness of the problem. Finally, criteria of viability of the  
$f(R)$-models are considered according to the various matter fields acting as sources.
\end{abstract}

\pacs{04.50.+h, 04.20.Ex, 04.20.Cv, 98.80.Jr}
\keywords{Alternative theories of gravity; metric approach; metric-affine approach; initial value formulation }

\maketitle
\section{Introduction}
The observed acceleration of  cosmic fluid   \cite{SN} has, up to now, no definitive 
explanation despite of the fact that  this discovery has dramatic
 implications not only for cosmology, but also for fundamental
physics.  According to the results of  {\em WMAP}, {\em PLANCK}
and  other experiments mapping anisotropies of the cosmic microwave
background, if General Relativity (GR) gives   the correct description
of gravitational interaction at cosmological level, then approximately 75\% of  cosmic energy content
is an exotic, invisible, and unclustered 
 form of {\em dark energy}.  Furthermore, in order to explain clustering of large scale structure and stability of galactic systems, we need almost 25\% of  another mysterious component called {\em dark matter}.  
 No final evidence for both these dark ingredients has been found at quantum level and   this status of art brought to the formulation of the so-called $\Lambda$CDM  model which  addressing the coarse-grained problem of cosmology does not satisfy the  issue to explain the fundamental nature of  dark energy and dark matter  \cite{book}.

An alternative   approach consists in dispensing  with
the mysterious dark energy and dark matter and asking for the fact that gravitational interaction could act in different way at large scales.
Historically, this approach
was pursued in explaining phenomena as  the precession of
Mercury's perihelion which is not due to an unseen mass but to Einstein's
modification of Newtonian gravity.  The paradigm consists in   not imposing {\it a priori} the form of  gravitational action but assuming that it could be dictated by data, observations and experiments at various scales. The simplest example from this point of view is  the so-called $f(R)$-gravity where ``modified'' gravity consists of modifications  of GR that could become important  at low curvatures,  late
in the matter era (infrared limit) or at high curvatures toward inflationary epoch  (ultraviolet limit). Specifically, the Einstein-Hilbert
action is linear in the Ricci scalar $R$ and could work only at local (e.g. Solar System) scales. Here $f(R)$ is a generic, non-linear function of $R$  \cite{CCT}.

In principle, an effective interaction action  can  contain several degrees of freedom:
tensor, vector, and scalar, massless or massive. In GR
only the massless spin~2 graviton propagates. When the
Einstein-Hilbert action is modified, other degrees of freedom
appear. The change $R \rightarrow f(R)$ in the action brings  to  a further massive scalar mode which
can drive both the today observed cosmic acceleration and the primordial inflation, depending on its mass. This
is analogous to the inflaton  field driving  the
accelerated expansion of the early universe, although at a much lower
energy scale.

If terms quadratic in
the Ricci and Riemann tensor, and possibly other curvature invariants,
are included in the gravitational Lagrangian, $ f
\left( R, R_{ab}R^{ab}, R_{abcd}R^{abcd}, \, ... \, \right)$,
massive  gravitons and vector degrees of freedom appear \cite{greci}. In the
following we restrict ourselves to
$f(R)$ theories of gravity   and focus on
their initial value problem \footnote{$f(R)$ gravity,
with an explicit coupling to matter to $R$ \cite{extraf}, can be
also  used as a valid alternative to  dark matter
\cite{DM,salzano}.}.  For  a more
comprehensive discussion on Extended Theories of Gravity, we refer the reader to \cite{book,review}. 

We adopt a conservative point of view and regard $f(R)$-gravity 
more as a toy model than the final theory of gravity, {\em
i.e.}, we consider these theories as  a proof of
principle that modifying gravity is a viable alternative to dark energy and dark matter.
However, we do not feel that one can claim that any of the $f(R)$ models
proposed thus far  is the
``correct'' one, or has exceptional support from the observational data.
While it is true that many $f(R)$ models pass all the available
experimental tests and fit the cosmological data, the same is true for
many dark energy models, and it is currently impossible to use
today available observational data to definitely discriminate
between most of them, and between dark energy and modified gravity
models.

Modifying gravity is risky: unwanted consequences may be  the violations of
 experimental limits on the
parametrized-post-Newtonian (PPN) parameters at terrestrial and
Solar System scales \cite{Will},
instabilities,
ghosts and, as in any newly proposed
theory, the Cauchy problem is a very delicate issue. 

$f(R)$ gravity has  a long history: its origins can be  traced back
to Weyl's  theory  in which a term quadratic in the Weyl tensor
was added to the Einstein-Hilbert Lagrangian \cite{Weyl}. Later, $f(R)$
gravity
received the attention  by many authors, including
Eddington, Bach, Lanczos, Schr\"odinger, and Buchdahl. In the 1960's and
1970's, it was found that  quadratic corrections to the Hilbert-Einstein action  were
necessary to improve the
renormalizability of GR \cite{renorma}, and in
1980 quadratic
corrections were found to fuel inflation without the need for scalar
fields \cite{Starobinsky80}. Non-linear corrections are also motivated
by string theories \cite{stringmotivations}. We refer the reader to
\cite{Schmidt} for a comprehensive historical review.

The prototype of $f(R)$ gravity  is the model
\be
f(R)=R+\alpha R^2-\mu^4/R \;,
\ee
where $\alpha$ is a mass parameter related to the so called {\it scalaron} \cite{Starobinsky80, Schmidt},  and $\mu $ is a mass scale of the order of the present value of the
Hubble parameter $\mu \sim H_0\sim 10^{-33}$~eV. Although ruled out
by its weak-field limit  \cite{CSE} and by
a violent instability \cite{DolgovKawasaki}, this model  gives the idea
underlying  modified gravity:  the $ 1/R$ correction
is negligible in comparison with $R^2$ at the high curvatures of the early
universe, and kicks in only as
$R\rightarrow 0$, late in the history of the universe.  GR is restored when the $R$-term is leading.

Several   models have been proposed   in  literature: here
we discuss only general  features of $f(R)$-gravity addressing, in particular, the Cauchy problem for metric and metric-affine formulations of these theories.

To be more specific, 
any theory of physics is "physically" viable if an appropriate
initial value problem is suitably formulated. This means that, starting from 
the assignment of suitable initial data, the
subsequent dynamical evolution of the physical system is uniquely
determined. In this case,
the problem is said {\it well-formulated}. However, also if the initial
value problem is  well-formulated, we need other 
properties that a viable theory has to satisfy. First of all,
small changes and perturbations in the initial data have to
produce small perturbations in the subsequent dynamics over all
the space-time where it is defined. This means that the theory
should be "stable" in order to be "predictive". Besides, changes
in the initial data region have to preserve the causal structure
of the theory. If both these requirements are satisfied, the
initial value problem of the theory is also {\it well-posed}.

 GR has a well-formulated and
well-posed initial value problem  but, as for other relativistic
field theories, we need initial value constraints and  gauge
choices in order to make Einstein's field equations suitable for a
correct formulation of the Cauchy problem. The consequence of this
well-posedness is that GR is a "stable" theory
with a robust causal structure where singularities can be
classified \cite{Synge,Wald}.

In this paper, we focus attention on the well--posedness of the initial value
problem of metric and metric-affine $f(R)$-theories of gravity. The aim is to prove that the Cauchy
problem is well posed in different and physically important cases.

The layout of the paper is the following. In Sec.II,  we give the generalities of metric formulation of $f(R)$-gravity stressing, in particular, how it can be recast in terms of scalar-tensor gravity (the so called O'Hanlon theory).  After we discuss the Cauchy problem showing that it is both well formulated and well posed.
The metric-affine formulation is discussed in Sec.III. In this case, the Cauchy problem has some subtleties that have to be faced with care. While the well formulation is in general always valid (at least for standard matter sources \cite{CV3}), the well--posedness may depends on the specific form of the function $f(R)$. The relevant cases of perfect fluid and scalar field as sources are considered in details. Sec.IV is devoted to the discussion of the results and the conclusions.

\section{Metric $f(R)\/$-gravity} 
\subsection{Generalities}
The action of  $f(R)\/$- gravity in metric formalism is given by
\begin{equation}\label{2.1.1}
{\cal A}\/(g)=\int{\sqrt{|g|}\left[ f\/(R) + {\cal L}_m \right]\,ds}
\end{equation}
where $f(R)$ is a real function of the scalar curvature $R\/(g) = g^{ij}R_{ij}\/$
(with $R_{ij}:= R^h_{\;\;ihj}\/$) and ${\cal L}_m\,ds\/$
is the material Lagrangian. The field equations are obtained by varying \eqref{2.1.1} with respect to the metric $g_{ij}\/$.  This yields 
\begin{equation}\label{2.1.2}
f'(R)R_{ij} - \frac{1}{2}f(R)g_{ij} = \Sigma_{ij} + \nabla_i\nabla_j f'(R) - g_{ij}g^{pq}\nabla_p\nabla_q f'(R)
\end{equation}
where ${\displaystyle\Sigma_{ij}:= -\frac{1}{\sqrt{|g|}}\frac{\delta{\cal L}_m}{\delta g^{ij}}\/}$ is the energy-momentum tensor of perfect-fluid matter. Eqs. \eqref{2.1.2} can be recast in Einstein-like form 
\begin{equation}\label{2.1.3}
\begin{split}
R_{ij} - \frac{1}{2}Rg_{ij} = \frac{1}{f'(R)}\left[ \Sigma_{ij} + \nabla_i\nabla_j\/f'(R) - g_{ij}g^{pq}\nabla_p\nabla_q\/f'(R) \right.\\
\left. - \frac{1}{2}\left(f'(R)R - f(R)\right)g_{ij} \right]
\end{split}
\end{equation}
$f(R)$-gravity can be re-interpreted as an O'Hanlon scalar-tensor theory, by introducing a suitable scalar field $\varphi$ which non-minimally couples with the gravity sector \cite{book,hanlon}. To see this point, let us take into account an O'Hanlon Lagrangian of the form 
\begin{equation}\label{2.1.4}
{\cal A}\/(g,\varphi)=\int{\sqrt{|g|}\left[\varphi\/R - V\/(\varphi) + {\cal L}_m \right]\,ds}
\end{equation}
where $V\/(\varphi)$ is the self--interaction potential. Field equations are derived by varying \eqref{2.1.4} with respect to both $g_{ij}$ and $\varphi$ which now represents a new dynamical variable. One obtains
\begin{subequations}\label{2.1.5}
\begin{equation}\label{2.1.5a}
R_{ij} - \frac{1}{2}Rg_{ij}= \frac{1}{\varphi}\left[ \Sigma_{ij} + \nabla_i\nabla_j\/\varphi - g_{ij}g^{pq}\nabla_p\nabla_q\/\varphi - \frac{1}{2}V\/(\varphi)g_{ij} \right]
\end{equation}
\begin{equation}\label{2.1.5b}
R - \frac{dV\/(\varphi)}{d\varphi} =0
\end{equation}
\end{subequations}
We notice that, replacing \eqref{2.1.5b} into  the trace of \eqref{2.1.5a}, we obtain the scalar equation 
\begin{equation}\label{2.1.6}
g^{pq}\nabla_p\nabla_q\/\varphi = \frac{1}{3}\left[ \varphi\/\frac{dV\/(\varphi)}{d\varphi} - 2V\/(\varphi) + \Sigma \right]
\end{equation}
System \eqref{2.1.5} is then equivalent to eqs. \eqref{2.1.5a} together with \eqref{2.1.6}. 
Given the function $f\/(R)$ in \eqref{2.1.1}, we shall suppose that its first derivative $f'\/(R)$ is invertible. In such a circumstance, it is easily seen that metric $f(R)$-theories of gravity can be mapped onto O'Hanlon theories and vice-versa. Indeed, defined the scalar field
\begin{subequations}\label{2.1.7}
\begin{equation}\label{2.1.7a}
\varphi:=f'\/(R)
\end{equation}
and the potential
\begin{equation}\label{2.1.7b}
V\/(\varphi):= f'\/\left[R\/(\varphi))R\/(\varphi) -f\/(R\/(\varphi)\right]
\end{equation}
\end{subequations}
It is  straightforward  to verify that, under the above  hypothesis $f''\/(R) \not = 0$, eq. \eqref{2.1.7a} expresses the inverse relation of \eqref{2.1.5b}, namely
\begin{equation}\label{2.1.8}
R - \frac{dV\/(\varphi)}{d\varphi} =0 \qquad\Longleftrightarrow\qquad \varphi:=f'\/(R)
\end{equation}
being the potential $V\/(\varphi)$ defined by \eqref{2.1.7b}. A direct comparison of eqs. \eqref{2.1.3} with eqs. \eqref{2.1.5a} shows then that solutions of \eqref{2.1.3} together with \eqref{2.1.7a} are also solutions of \eqref{2.1.5} and viceversa.
As a final remark, we recall that in O'Hanlon theory the standard conservation laws $\nabla^i\Sigma_{ij}=0\/$ hold. An explicit proof of the vanishing of the covariant divergence of the energy-momentum tensor in modified theories of gravity can be found in \cite{Koivisto}.

\subsection{The Cauchy problem for O'Hanlon gravity}

Taking into account the above  dynamical equivalence, the Cauchy problem for  $f(R)$- gravity can be defined as the Cauchy problem for the corresponding O'Hanlon theory. 
In this perspective, we discuss  the Cauchy problem for the  O'Hanlon gravity. Let us show the well-posedness of the Cauchy problem for  system \eqref{2.1.5a} and \eqref{2.1.6} in vacuo. As we shall see, the same conclusions hold in presence of matter sources satisfying the standard conservation laws $\nabla^i\Sigma_{ij}=0\/$. To  this aim, we  use generalized harmonic coordinates, given by the conditions
\begin{equation}\label{2.2.1}
F^i_{\varphi}:= F^i - H^i =0 \qquad {\rm with}\qquad F^i :=g^{pq}\Gamma^i_{pq}, \quad H^i := \frac{1}{\varphi}\nabla^i\/\varphi
\end{equation} 
The generalized harmonic gauge \eqref{2.2.1} is a particular case of the one introduced in \cite{Salgado} to prove the well-posedness of the Cauchy problem for more general scalar-tensor theories of gravity. As we shall see, the gauge \eqref{2.2.1} allows us to develop a second order analysis very similar to the one used in GR \cite{yvonne4}. 
We rewrite eqs. \eqref{2.1.5a} in the form
\begin{equation}\label{2.2.2}
R_{ij} = \frac{1}{\varphi}\left[T_{ij} - \frac{1}{2}T\/g_{ij}\right]
\end{equation}
where
\begin{equation}\label{2.2.3}
T_{ij}:= \nabla_i\nabla_j\/\varphi - g_{ij}g^{pq}\nabla_p\nabla_q\/\varphi - \frac{1}{2}V\/(\varphi)g_{ij}
\end{equation}
plays the role of effective energy--momentum tensor. The Ricci tensor can be expressed as \cite{yvonne4}
\begin{equation}\label{2.2.4}
R_{ij} = R_{ij}^\varphi + \frac{1}{2}\left[ g_{ip}\partial_j\left( F^p_\varphi + H^p \right) + g_{jp}\partial_i\left( F^p_\varphi + H^p \right)\right]
\end{equation}
with
\begin{equation}\label{2.2.5}
R_{ij}^\varphi := - \frac{1}{2}g^{pq}\partial^2_{pq}\/g_{ij} + A_{ij}\/(g,\partial g)
\end{equation}
where only  first order derivatives  appear in the functions $A_{ij}$. Assuming $F^i_\varphi =0$ and taking the expression of $H^i$ into account, we obtain the following representation
\begin{equation}\label{2.2.6}
R_{ij} = - \frac{1}{2}g^{pq}\partial^2_{pq}\/g_{ij} + \frac{1}{\varphi}\partial^2_{ij}\/\varphi + B_{ij}\/(g,\varphi,\partial g,\partial\varphi)
\end{equation}
where the functions $B_{ij}$ depend on the metric $g$, the scalar field $\varphi$ and their first order derivatives. Analogously, using eq. \eqref{2.1.6} to replace all terms depending on the divergence $g^{pq}\nabla_p\nabla_q\/\varphi$, the right hand side of \eqref{2.2.2} can be expressed as 
\begin{equation}\label{2.2.7}
\frac{1}{\varphi}\left[T_{ij} - \frac{1}{2}T\/g_{ij}\right] = \frac{1}{\varphi}\partial^2_{ij}\/\varphi + C_{ij}\/(g,\varphi,\partial g,\partial\varphi)
\end{equation}
Again, in the functions $C_{ij}$, only first order derivatives are involved. A direct comparison of eqs. \eqref{2.2.6} with eqs. \eqref{2.2.7} shows  that, in the considered gauge, eqs. \eqref{2.2.1} assume the form  
\begin{equation}\label{2.2.8}
g^{pq}\partial^2_{pq}\/g_{ij} = D_{ij}\/(g,\varphi,\partial g,\partial\varphi)
\end{equation}
Eqs. \eqref{2.2.8}, together with eq. \eqref{2.2.6}, form a quasi-diagonal, quasi-linear second-order system of partial differential equations, for which well known theorems by Leray \cite{yvonne4,Leray,Wald} hold. Given initial data on a space--like surface, the associated Cauchy problem is then well-posed in suitable Sobolev spaces \cite{yvonne4}. Of course, the initial data have to satisfy the gauge conditions $F^i_{\varphi}=0$ as well as the Hamiltonian and momentum constraints
\begin{equation}\label{2.2.9}
G^{0i}=\frac{1}{\varphi}\/T^{0i} \quad i=0,\ldots,3
\end{equation}
on the initial space-like surface. In connection with this, we notice that, from eq. \eqref{2.1.6}, we can derive the expression of the second partial derivative $\partial^2_0\/\varphi$ and replace it in the right hand side of \eqref{2.2.9}, so obtaining constraints involving no higher than first order partial derivatives with respect to the time variable $x^0\/$. To conclude, we have to prove that the gauge conditions $F^i_\varphi =0$ are preserved in a neighborhood of the initial space-like surface. To this end, we first verify that the divergence of the Einstein-like equations \eqref{2.1.5a} vanishes, namely
\begin{equation}\label{2.2.10}
\nabla^i\/\left(\varphi\/G_{ij} - T_{ij}\right) =0
\end{equation}
A straightforward calculation yields
\begin{equation}\label{2.2.11}
\begin{split}
\nabla^i\/\left(\varphi\/G_{ij} - T_{ij}\right) = \left(\nabla^i\/\varphi\right)\/R_{ij} -\frac{1}{2}\varphi_j\left( R - \frac{dV}{d\varphi} \right) 
+\varphi\nabla^i\/G_{ij} \\
- \left( \nabla^i\nabla_i\nabla_j 
- \nabla_j\nabla^i\nabla_i \right)\varphi
\end{split}
\end{equation}
By definition, the Einstein and Ricci tensors satisfy the identities $\nabla^i\/G_{ij}=0\/$ and $\left(\nabla^i\/\varphi\right)\/R_{ij}=\left( \nabla^i\nabla_i\nabla_j - \nabla_j\nabla^i\nabla_i \right)\varphi$. On the other hand, 
${\displaystyle R - \frac{dV}{d\varphi}=0}$ is assured by  field equations \eqref{2.1.5b}. Therefore,  identities \eqref{2.2.10} follow. If now $g_{ij}$ and $\varphi$ are the solutions of  reduced Einstein-like equations 
\eqref{2.2.8} and  field equation \eqref{2.1.6}, one has
\begin{equation}\label{2.2.12}
\varphi\/G^{ij} - T^{ij} = - \frac{\varphi}{2}\/\left( g^{ip}\partial_p\/F^j_\varphi + g^{jp}\partial_p\/F^i_\varphi - g^{ij}\partial_p\/F^p_\varphi \right)
\end{equation}
Identity \eqref{2.2.10} shows then that the functions $F^i_\varphi$ satisfy necessarily the linear homogeneous system of wave equations
\begin{equation}\label{2.2.13}
g^{pq}\partial^2_{pq}\/F^i_\varphi + E^{iq}_p\/\partial_q\/F^p_\varphi =0
\end{equation}
where $E^{iq}_p\/$ are known functions on the space-time. Since the constraints \eqref{2.2.9} amount to the condition $\partial_0\/F^i_\varphi$ \cite{yvonne4} on the initial space-like surface, a well known uniqueness theorem for differential systems like \eqref{2.2.13} \cite{yvonne4} assures that $F^i_\varphi=0$ in the region where solutions of \eqref{2.2.8} and \eqref{2.1.6} exist.

As mentioned above, the well-posedness of the Cauchy problem can be proved also in presence of coupling with standard matter sources, such as electromagnetic or Yang-Mills fields, (charged) perfect fluid, (charged) dust, Klein-Gordon scalar fields. When this is the case, eqs. \eqref{2.1.6} and \eqref{2.2.8} have to be coupled with the matter field equations. Applying the same arguments developed for GR  
\cite{yvonne4,yvonne2,yvonne,yvonne3}, it is easily seen that, in the generalized harmonic gauge \eqref{2.2.1}, the matter field equations together with eqs. \eqref{2.1.6} and \eqref{2.2.8} form a Leray hyperbolic and causal differential system  admitting a well-posed Cauchy problem. In addition to the well-known results by Bruhat's, the key point is that the field equations of matter field imply the standard conservation laws $\nabla^i\/\Sigma_{ij}=0$. This fact allows to verify the validity of eqs. \eqref{2.2.10} in presence of matter too.

\section{Metric-affine $f(R)\/$-gravity}
\subsection{Generalities}

In the metric-affine approach  to $f(R)$-gravity, the
(gravitational)  dynamical fields are a pseudo-Riemannian metric $g\/$ and a linear
connection $\Gamma\/$ on the space-time manifold $M\/$. 
Specifically, in the so-called Palatini approach, the connection
$\Gamma\/$ is torsionless and it is not requested to be
metric-compatible. In the approach with torsion, the
dynamical connection $\Gamma$ is forced to be metric but it has
torsion different from zero.
The field equations are derived from an action functional of the form
\begin{equation}\label{2.1}
{\cal A}\/(g,\Gamma)=\int{\left(\sqrt{|g|}f\/(R) + {\cal L}_m\right)\,ds}
\end{equation}
where  now $R\/(g,\Gamma) = g^{ij}R_{ij}\/$
(with $R_{ij}:= R^h_{\;\;ihj}\/$)  is the scalar curvature
associated with the connection $\Gamma\/$.
Throughout the paper we shall assume that the material Lagrangian is independent of the dynamical connection. In this case, the field equations are
\begin{subequations}\label{2.2}
\begin{equation}\label{2.2a}
f'\/(R)R_{ij} - \frac{1}{2}f\/(R)g_{ij}=\Sigma_{ij}\,,
\end{equation}
\begin{equation}\label{2.2b}
T_{ij}^{\;\;\;h} = - \frac{1}{2f'\/(R)}\de{f'\/(R)}/de{x^p}\/\left(\delta^p_i\delta^h_j - \delta^p_j\delta^h_i\right)
\end{equation}
\end{subequations}
for $f(R)$-gravity with torsion \cite{CCSV1,CCSV2,CCSV3}, and
\begin{subequations}\label{2.3}
\begin{equation}\label{2.3a}
f'\/(R)R_{ij} - \frac{1}{2}f\/(R)g_{ij}=\Sigma_{ij}\,,
\end{equation}
\begin{equation}\label{2.3b}
\nabla_k\/(f'(R)g_{ij})=0\,,
\end{equation}
\end{subequations}
for $f(R)$-gravity in the  Palatini approach \cite{francaviglia1,francaviglia2,Olmo}. 
Considering the trace of Eqs.
\eqref{2.2a} and \eqref{2.3a}, we get the relation 
\begin{equation}\label{2.4}
f'\/(R)R -2f\/(R) = \Sigma
\end{equation}
which links the curvature scalar $R\/$ with the trace of the stress-energy tensor
$\Sigma:=g^{ij}\Sigma_{ij}\/$.
From now on, we shall assume that the relation \eqref{2.4} is
invertible that is $\Sigma\not=\const\/$ (this implies,
for example, $f(R)\/ \neq \alpha R^2\/$ which is only
compatible with $\Sigma=0\/$) . Under these hypotheses, the
curvature scalar $R\/$ can be expressed as a suitable function of
$\Sigma\/$, namely
\begin{equation}\label{2.5}
R=F(\Sigma)\,.
\end{equation}
If $\Sigma=\const$, GR plus the cosmological constant is recovered \cite{CCSV1}.
Defining the scalar field
\begin{equation}\label{2.6}
\varphi:=f'(F(\Sigma))
\end{equation}
we can put the Einstein-like field equations for both {\it \`a
la\/} Palatini  and with torsion theories in the same form
\cite{CCSV1,Olmo}, that is
\begin{equation}\label{2.7}
\begin{split}
\tilde{R}_{ij} -\frac{1}{2}\tilde{R}g_{ij}= \frac{1}{\varphi}\Sigma_{ij}
+ \frac{1}{\varphi^2}\left( - \frac{3}{2}\de\varphi/de{x^i}\de\varphi/de{x^j}
+ \varphi\tilde{\nabla}_{j}\de\varphi/de{x^i} + \frac{3}{4}\de\varphi/de{x^h}\de\varphi/de{x^k}g^{hk}g_{ij} \right. \\
\left. - \varphi\tilde{\nabla}^h\de\varphi/de{x^h}g_{ij} -
V\/(\varphi)g_{ij} \right)\,
\end{split}
\end{equation}
where we have introduced the effective potential
\bigskip\noindent
\begin{equation}\label{2.8}
V\/(\varphi):= \frac{1}{4}\left[ \varphi
F^{-1}\/((f')^{-1}\/(\varphi)) +
\varphi^2\/(f')^{-1}\/(\varphi)\right]\,
\end{equation}
for the scalar field $\varphi\/$. In Eq. \eqref{2.7}, $\tilde{R}_{ij}\/$, $\tilde{R}\/$ and $\tilde\nabla\/$ denote respectively the Ricci tensor, the scalar curvature and the covariant derivative associated with the Levi--Civita connection of the dynamical metric $g_{ij}$.
Therefore, if the dynamical connection $\Gamma\/$ is not coupled
with matter, both the theories (with torsion and Palatini--like)
generate identical Einstein--like field equations. 

In addition to this, it has been shown \cite{CCSV1} that the Einstein-like equations \eqref{2.7} (together with eqs.\eqref{2.6}) are deducible from a scalar-tensor theory with Brans-Dicke parameter $\omega_0=-3/2\/$. To see this point, we recall that the action functional of a (purely metric) scalar-tensor theory is given by
\begin{equation}\label{00001}
{\cal A}\/(g,\varphi)=\int{\left[\sqrt{|g|}\left(\varphi\tilde{R} -\frac{\omega_0}{\varphi}\varphi_i\varphi^i - U\/(\varphi) \right)+ {\cal L}_m\right]\,ds}
\end{equation}
where $\varphi\/$ is the scalar field, ${\displaystyle \varphi_i := \de\varphi/de{x^i}\/}$ and $U\/(\varphi)\/$ is the potential of $\varphi\/$. The matter Lagrangian ${\cal L}_m\/(g_{ij},\psi)\/$ is a function of  metric and some matter fields $\psi\/$; $\omega_0\/$ is the so called Brans--Dicke parameter.
 
The variation of \eqref{00001} with respect to the metric and the scalar field yields the field equations
\begin{equation}\label{0000.2}
\tilde{R}_{ij} -\frac{1}{2}\tilde{R}g_{ij}= \frac{1}{\varphi}\Sigma_{ij} + \frac{\omega_0}{\varphi^2}\left( \varphi_i\varphi_j  - \frac{1}{2}\varphi_h\varphi^h\/g_{ij} \right) 
+ \frac{1}{\varphi}\left( \tilde{\nabla}_{j}\varphi_i - \tilde{\nabla}_h\varphi^h\/g_{ij} \right) - \frac{U}{2\varphi}g_{ij} 
\end{equation} 
and
\begin{equation}\label{0000.3}
\frac{2\omega_0}{\varphi}\tilde{\nabla}_h\varphi^h + \tilde{R} - \frac{\omega_0}{\varphi^2}\varphi_h\varphi^h - U' =0
\end{equation} 
where ${\displaystyle \Sigma_{ij}:= - \frac{1}{\sqrt{|g|}}\frac{\delta{\cal L}_m}{\delta g^{ij}}\/}$ and ${\displaystyle U' :=\frac{dU}{d\varphi}\/}$.
Taking the trace of eq. \eqref{0000.2} and using it to replace $\tilde R\/$ in eq. \eqref{0000.3}, one obtains the equation
\begin{equation}\label{0000.4}
\left( 2\omega_0 + 3 \right)\/\tilde{\nabla}_h\varphi^h = \Sigma + \varphi U' -2U
\end{equation} 
By a direct comparison, it is immediately seen that for 
${\displaystyle \omega_0 :=-\frac{3}{2}\/}$ and ${\displaystyle U\/(\varphi) :=\frac{2}{\varphi}V\/(\varphi)\/}$ (where $V\/(\varphi)\/$ is defined in eq. \eqref{2.8})  
eqs. \eqref{0000.2} become formally identical to the Einstein-like equations \eqref{2.7} 
for a metric--affine $f(R)\/$ theory. Moreover, in such a circumstance, eq. \eqref{0000.4} reduces to the algebraic equation
\begin{equation}\label{0000.5}
\Sigma + 2V'\/(\varphi) -\frac{6}{\varphi}V\/(\varphi) =0
\end{equation}   
relating the matter trace $\Sigma\/$ to the scalar field $\varphi\/$, exactly as it happens for metric-affine $f(R)\/$-gravity. In particular, it is a straightforward matter to verify that (under the condition $f''\not= 0\/$ 
\cite{CCSV1}) eq. \eqref{0000.5} expresses exactly the inverse relation of \eqref{2.6}, namely
\begin{equation}\label{0000.6}
\Sigma + 2V'\/(\varphi) -\frac{6}{\varphi}V\/(\varphi) =0 \quad \Longleftrightarrow \quad \Sigma=F^{-1}\/((f')^{-1}\/(\varphi))
\end{equation}
being $F^{-1}\/(X) = f'\/(X)X - 2f\/(X)\/$. When the dynamical connection does not couple with matter, metric-affine $f(R)\/$-gravity (with torsion or Palatini-like) are then dynamically equivalent to scalar-tensor theories with a Brans--Dicke parameter ${\displaystyle \omega_0 =-\frac{3}{2}\/}$.

Moreover, we notice that  field equations \eqref{2.7} can be simplified by passing from the Jordan to the Einstein frame.
Indeed, performing the conformal transformation $\bar{g}_{ij}=\varphi\/g_{ij}$, eqs. \eqref{2.7} assume the simpler
form (see for example \cite{CCSV1,Olmo})
\begin{equation}\label{3.1.1}
\bar{R}_{ij} - \frac{1}{2}\bar{R}\bar{g}_{ij} =
\frac{1}{\varphi}\Sigma_{ij} -
\frac{1}{\varphi^3}V\/(\varphi)\bar{g}_{ij}
\end{equation}
where $\bar{R}_{ij}\/$ and $\bar{R}\/$ are respectively the Ricci
tensor and the curvature scalar derived from the conformal metric
$\bar{g}_{ij}\/$. 

In connection with this fact, we recall that the conservation laws existing in the Jordan and in the Einstein frame, as well as their relations, are given by the following \cite{CV1,CV3}
\begin{Proposition}\label{Pro3.1}
Equations \eqref{2.7}, \eqref{2.8} and \eqref{0000.5} imply the standard conservation laws $\tilde{\nabla}^j\Sigma_{ij}=0\/$.
\end{Proposition}
\begin{Proposition}\label{Pro3.2}
The condition $\tilde{\nabla}^j\Sigma_{ij}=0\/$ is equivalent to the condition $\bar{\nabla}^jT_{ij}=0\/$, where 
${\displaystyle T_{ij}:=\frac{1}{\varphi}\Sigma_{ij} - \frac{1}{\varphi^3}V\/(\varphi)\bar{g}_{ij}\/}$ and $\bar\nabla\/$ denotes the covariant derivative associated to the metric $\bar{g}_{ij}\/$.
\end{Proposition}

\subsection{The Cauchy problem in metric-affine formulation}
As mentioned above, if the trace of the stress-energy tensor $\Sigma =\const\/$ both Palatini-like and with torsion theories reduce   to GR  with cosmological constant. Therefore, in such a circumstance, the Cauchy problem is well-formulated and well-posed \cite{yvonne4,yvonne,Wald}. For example, this is the case in vacuo and in presence of electromagnetic (or also Yang-Mills) fields (if $f(R) \not = \alpha\/R^2\/$). 

On the contrary, if $\Sigma \not = \const\/$, the situation is more complicated. In this case, the theory does not longer reduce to GR and so the well-formulation and well-posedness of the Cauchy problem is not automatically assured by the classical Bruhat results. 

In analogy with the purely metric case, matter sources could be again considered for Cauchy problem by using the dynamical equivalence with scalar-tensor theories  adopting a Brans-Dicke parameter ${\displaystyle \omega_0 =-\frac{3}{2}\/}$. Unfortunately, here a difficulty occurs. The point is that the d'Alembertian $g^{pq}\nabla_p\nabla_q\varphi\/$ disappears from eq. \eqref{0000.4}. It follows the impossibility of deriving the expression of the d'Alembertian $g^{pq}\nabla_p\nabla_q\varphi\/$ as a function of the dynamical variables and their derivatives up to the first-order. In other words, we cannot eliminate the second-order derivatives of the scalar field $\varphi$ from the Einstein-like equations \eqref{0000.2}. This fact led some authors \cite{Faraoni} to the conclusion that Palatini $f(R)$-gravity has an ill-formulated Cauchy problem, so claiming a sort of no-go theorem for these theories.

In the subsequent discussion, we shall prove that the conclusions achieved in \cite{Faraoni} are not always true. Indeed, using the conformal transformation technique to pass from the Jordan to the Einstein frame, we shall give sufficient conditions ensuring the well-posedness of the Cauchy problem for metric-affine $f(R)$-gravity coupled with a perfect fluid \cite{CV1,CV2} or a Klein-Gordon scalar field \cite{CV5}. The stated conditions will result as suitable requirements imposed on the function $f(R)$, so representing a sort of  selection rules for viable $f(R)$ models. Moreover, we shall show that the set of viable $f(R)$ models is not empty since the function $f(R)=R+\alpha\/R^2\/$ satisfies the above mentioned conditions.

\subsection{The Cauchy problem in presence of a perfect fluid}

Let us discuss the Cauchy problem for metric-affine $f(R)\/$-gravity coupled with a perfect fluid. It is possible to  show that, passing to the Einstein frame through a conformal transformation, the analysis of the initial value problem can be carried out by following the same arguments developed in \cite{yvonne4,yvonne2,yvonne} for GR.
Let us consider the metric $g_{ij}\/$ of signature $(- + + +)$ in the Jordan frame and a perfect fluid with stress-energy tensor of the form
\begin{subequations}\label{4.1}
\begin{equation}\label{4.1a}
\Sigma_{ij}=(\rho + p)\,U_iU_j + p\,g_{ij}
\end{equation}
The corresponding matter field equations are 
\begin{equation}\label{4.1b}
\tilde\nabla_j\Sigma^{ij}=0
\end{equation}
\end{subequations}
In eqs. \eqref{4.1}, the scalars $\rho$ and $p$ denote respectively the matter-energy density and the pressure of the fluid, while $U_i$ 
indicate the four velocity of the fluid satisfying the condition $g^{ij}U_iU_j =-1\/$.
If we perform the conformal transformation $\bar{g}_{ij}=\varphi g_{ij}\/$, we can express the field equations in the Einstein frame as
\begin{subequations}\label{4.2}
\begin{equation}\label{4.2a}
\bar{R}_{ij} - \frac{1}{2}\bar{R}\bar{g}_{ij} = T_{ij}
\end{equation}
and 
\begin{equation}\label{4.2b}
\bar{\nabla}_j T^{ij}=0
\end{equation}
\end{subequations}
where
\begin{equation}\label{4.3}
T_{ij}=\frac{1}{\varphi}(\rho + p)\,U_iU_j + \left( \frac{p}{\varphi^2} - \frac{V(\varphi)}{\varphi^3} \right)\,\bar{g}_{ij}
\end{equation}
plays the role of the effective stress-energy tensor.
Due to Proposition \ref{Pro3.2}, eqs. \eqref{4.2b} are equivalent to eqs. \eqref{4.1b}. This point is crucial in our discussion, because it allows us to apply to the present case the results achieved in \cite{yvonne4,yvonne2,yvonne}. 
For the sake of simplicity, we shall suppose that the scalar field $\varphi$ is positive, that is $\varphi > 0\/$. The opposite case $\varphi <0\/$ differs from the former only for some technical aspects and it will be briefly outlined after. Of course, using the above conformal transformation, it is implicitly assumed that $\varphi \not = 0\/$ at least in a neighbourhood of the initial space-like surface.

Under these assumptions, the four velocity of the fluid in the Einstein frame can be expressed as $\bar{U}_i = \sqrt{\varphi}U_i\/$, so that the stress-energy tensor \eqref{4.3} can be rewritten in terms of the four velocity $\bar{U}_i\/$ as
\begin{equation}\label{4.4}
T_{ij}=\frac{1}{\varphi^2}(\rho + p)\,\bar{U}_i\bar{U}_j + \left( \frac{p}{\varphi^2} - \frac{V(\varphi)}{\varphi^3} \right)\,\bar{g}_{ij}
\end{equation}
After that, let us introduce the effective mass-energy density
\begin{subequations}\label{4.5}
\begin{equation}\label{4.5a}
\bar{\rho}:= \frac{\rho}{\varphi^2} + \frac{V(\varphi)}{\varphi^3}
\end{equation}
and the effective pressure
\begin{equation}\label{4.5b}
\bar{p}:= \frac{p}{\varphi^2} - \frac{V(\varphi)}{\varphi^3}
\end{equation}
\end{subequations}
In view of eqs.	\eqref{4.5}, the stress--energy tensor \eqref{4.4} assumes the usual form
\begin{equation}\label{4.6}
T_{ij}=\left( \bar\rho + \bar{p} \right)\,\bar{U}_i\bar{U}_j + \bar{p}\,\bar{g}_{ij}
\end{equation}
It is worth noticing that, starting from an equation of state of the form $\rho=\rho(p)\/$ and assuming that  relation \eqref{4.5b} is invertible $(p=p(\bar{p}))$, by composition with eq. \eqref{4.5a}, we derive an effective equation of state $\bar{\rho}=\bar{\rho}(\bar{p})\/$. In addition, we recall that the explicit expression of the scalar field $\varphi\/$ as well as of the potential $V(\varphi)\/$ are directly related with the particular form of the function $f(R)\/$. Then, the requirement of invertibility of the relation \eqref{4.5b} together with the condition $\varphi >0\/$ (or, equivalently, $\varphi <0\/$) become criteria for the viability of the functions $f(R)\/$. In other words, they provide us with   selection rules for the admissible functions $f(R)\/$ (see also \cite{olmonew}).

At this point, the discussion of the Cauchy problem can proceed following step by step  Bruhat's arguments   \cite{yvonne4,yvonne2,yvonne}.  Bruhat's results are well known in literature, so we do not repeat the analysis here. We only recall that the Cauchy problem for the system of differential equations \eqref{4.2}, with stress-energy tensor given by eq. \eqref{4.6} and equation of state $\bar{\rho}=\bar{\rho}(\bar{p})\/$, is well-posed if the condition
\begin{equation}\label{4.7}
\frac{d\bar{\rho}}{d\bar{p}}\geq 1
\end{equation}
is satisfied. It is worth  noticing that, in order to check the requirement \eqref{4.7}, it is not necessary to invert explicitly the relation \eqref{4.5b}, but  simply we can verify the relation
\begin{equation}\label{4.8}
\frac{d\bar{\rho}}{d\bar{p}}=\frac{d\bar{\rho}/dp}{d\bar{p}/dp}\geq 1
\end{equation}
directly from  expressions \eqref{4.5} and  equation of state $\rho=\rho(p)\/$. Once again,  condition \eqref{4.8} depends on the  expressions of $\varphi$ and $V(\varphi)\/$, therefore it is strictly related to the  form of the function $f(R)\/$. Thus, condition \eqref{4.8} represents a further criterion for the admissibility of  $f(R)\/$-models.

For the sake of completeness, we outline the case $\varphi <0\/$. If we still suppose that the signature of the metric in the Jordan frame is $(-+++)\/$, the signature of the conformal metric is $(+---)\/$ and the four velocity of the fluid in the Einstein frame is then $\bar{U}_i = \sqrt{-\varphi}U_i\/$.
The effective stress-energy tensor is given now by
\begin{equation}\label{4.9}
T_{ij}=-\frac{1}{\varphi^2}(\rho + p)\,\bar{U}_i\bar{U}_j + \left( \frac{p}{\varphi^2} - \frac{V(\varphi)}{\varphi^3} \right)\,\bar{g}_{ij}= \left( \bar\rho + \bar{p} \right)\,\bar{U}_i\bar{U}_j - \bar{p}\,\bar{g}_{ij}
\end{equation}
where we have introduced the quantities
\begin{subequations}\label{4.10}
\begin{equation}\label{4.10a}
\bar{\rho}:= -\frac{\rho}{\varphi^2} - \frac{V(\varphi)}{\varphi^3}
\end{equation}
and
\begin{equation}\label{4.10b}
\bar{p}:= -\frac{p}{\varphi^2} + \frac{V(\varphi)}{\varphi^3}
\end{equation}
\end{subequations}
representing, as above, the effective mass-energy and the effective pressure. At this point, everything proceeds again as in \cite{yvonne4,yvonne2,yvonne}, except for a technical aspect: now the quantity
 ${\displaystyle r:=\bar{\rho}+\bar{p}=-\frac{\rho + p}{\varphi^2}\/}$ is negative (if, as usual, $\rho\/$ and $p\/$ are assumed positive). In this regard, the reader can easily verify that, with the choice $\log(-f^{-2}r)$ instead of $\log(f^{-2}r)$ as in \cite{yvonne4,yvonne2,yvonne}, the Bruhat's arguments apply equally well. 

Let us now give a simple and illustrative example of the above results.
Let us take into account the $f(R)= R+ \alpha{R^2}$ model coupled with dust (for, example, let us consider a cosmological model). The matter stress-energy tensor
in the Jordan frame is $\Sigma_{ij} = \rho U_i U_j$. 
The trace of the Einstein-like equations \eqref{2.2a} yields the relation
\begin{equation}\label{5.1} 
(1+ 2\alpha R)R -2R - 2\alpha R^2 = -\rho \qquad
\longleftrightarrow \qquad R=\rho
\end{equation} 
The scalar field \eqref{2.6} is given by  
\begin{equation}\label{5.2} 
\varphi(\rho) = f'(R(\rho))= 1+2\alpha\rho
\end{equation}
For small values of the density $\rho<<1\/$ (for example, the present cosmological density) and values of $|\alpha|\/$ not comparable with $1/\rho\/$, we can reasonably suppose $\varphi>0\/$, independently of the sign of the parameter $\alpha\/$. Let us calculate now the potential \eqref{2.8}
\begin{equation}\label{5.3}
V(\varphi)= \frac{1}{4}\left[ \varphi F^{-1}((f')^{-1}(\varphi)) +
\varphi^2(f')^{-1}(\varphi)\right]
\end{equation}
Since $(f')^{-1}(\varphi)=\rho$, from eq. \eqref{5.2} we have
\begin{equation}\label{5.4}
\frac{1}{4}\varphi^2(f')^{-1}(\varphi)=\frac{1}{4}(1+2\alpha\rho)^2\rho\,,
\end{equation}
also, considering the relation $F^{-1}(Y)=f'(Y)Y -2f(Y)$, it is
\begin{equation}\label{5.5}
\frac{1}{4}F^{-1}((f')^{-1}(\varphi))=\frac{1}{4}F^{-1}(\rho)=-\rho\,.
\end{equation}
and
\begin{equation}\label{5.6}
\frac{1}{4}\varphi F^{-1}((f')^{-1}(\varphi))=
-\frac{(1+2\alpha\rho)\rho}{4}\,,
\end{equation}
We can conclude that
\begin{equation}\label{5.7}
V(\varphi(\rho))= \frac{\alpha\rho^2(1+2\alpha\rho)}{2}
\end{equation}
In the Einstein frame, the stess-energy tensor is expressed as (comparing with eq. \eqref{4.4})
\begin{equation}\label{5.8}
T_{ij}= \frac{\rho}{\varphi^2}\bar{U}_i\bar{U}_j -\frac{V(\varphi)}{\varphi^3}\bar{g}_{ij}
\end{equation}
The latter can be seen as the stress-energy tensor of a perfect fluid with density and pressure given respectively by
\begin{subequations}\label{5.9}
\begin{equation}\label{5.9a}
\bar{\rho}:= \frac{\rho}{\varphi^2} + \frac{V(\varphi)}{\varphi^3}=\frac{2\rho + \alpha\rho^2}{2(1+2\alpha\rho)^2}
\end{equation}
and
\begin{equation}\label{5.9b}
\bar{p}:= - \frac{V(\varphi)}{\varphi^3}=-\frac{\alpha\rho^2}{2(1+2\alpha\rho)^2}
\end{equation}
\end{subequations}
Under the above  assumptions, it is easily seen that  function \eqref{5.9b} is invertible. For $\rho>0\/$, we have
\begin{equation}\label{5.10}
\frac{d\bar{p}}{d\rho}= -\frac{4\alpha\rho}{4\/(1+2\alpha\rho)^3}\not= 0
\end{equation}
Moreover, we have
\begin{equation}\label{5.11}
\frac{d\bar{\rho}}{d\rho}=\frac{4-4\alpha\rho}{4\/(1+2\alpha\rho)^3}
\end{equation}
so that 
\begin{equation}\label{5.12}
\frac{d\bar{\rho}}{d\bar{p}}=\frac{d\bar{\rho}/dp}{d\bar{p}/dp}=\frac{-1+\alpha\rho}{\alpha\rho}\geq 1 \quad\Longleftrightarrow\quad \alpha<0
\end{equation}
The conclusion is that  the model $f(R)=R + \alpha\/R^2\/$, with $\alpha < 0$, has  a well-posed Cauchy problem when coupled with dust. This is in agreement with results in \cite{barrow}.

\subsection{The Cauchy problem in presence of a scalar field}

Let us consider now the Cauchy problem for metric-affine 
$f(R)\/$-gravity coupled with a Klein-Gordon scalar field. 
As above, we shall state sufficient conditions ensuring
the well-posedness of the problem. 
Let $\psi$ be a Klein-Gordon
scalar field with self-interacting potential
${\displaystyle U(\psi)=\frac{1}{2} m^2\psi^2} $. The associated
stress-energy tensor is expressed as
\begin{equation}\label{3.1}
\Sigma_{ij}= \de\psi/de{x^i}\de\psi/de{x^j}
-\frac{1}{2}g^{ij}\left(\de\psi/de{x^p}\de\psi/de{x^q}g^{pq} +
m^2\psi^2\right)
\end{equation}
The corresponding Klein-Gordon equation is given by
\begin{equation}\label{3.2}
\tilde\nabla_j\de\psi/de{x^i}g^{ij}=m^2\psi
\end{equation}
where $\tilde{\nabla}\/$ denotes the Levi-Civita covariant derivative induced by the metric $g_{ij}\/$. The trace of  tensor \eqref{3.1} is 
\begin{equation}\label{3.3}
\Sigma :=\Sigma_{ij}g^{ij} = -\de\psi/de{x^p}\de\psi/de{x^q}g^{pq} -2m^2\psi^2\,.
\end{equation}
We notice that  trace \eqref{3.3} depends explicitly on the metric tensor $g_{ij}$. In view of this fact, we cannot apply the conformal transformation directly to the field equations \eqref{2.7}, being the scalar field $\varphi$ defined by \eqref{2.6}. Indeed, if we proceed in this way, both the metric $g_{ij}$ and $\bar{g}_{ij}$ would appear in the conformally transformed equations \eqref{3.1.1}. The idea to overcome this difficulty is using the already mentioned dynamical equivalence with ${\displaystyle \omega_0=-\frac{3}{2}}$ Brans-Dicke gravity. In other words, we discuss the Cauchy problem for a ${\displaystyle \omega_0=-\frac{3}{2}}$ Brans-Dicke theory coupled with the given Klein-Gordon field $\psi\/$.  
The field equations of such a theory are the Einstein-like equations \eqref{2.7}, the equation \eqref{0000.5} and the Klein--Gordon equation \eqref{3.2}. 
We perform the conformal transformation $\bar{g}_{ij}=\varphi\/g_{ij}\/$. The scalar field $\varphi$ is not longer defined by \eqref{2.6}, but it is now a dynamical variable related to the trace $\Sigma$ through eq. \eqref{0000.5}. This is an important difference with respect to the perfect fluid case discussed above. Anyway, after conformal transformation, the Einstein-like equations \eqref{2.7} assume the simpler form \eqref{3.1.1}.  
Moreover, using the relation 
\begin{equation}\label{3.5.0}
\bar{\Gamma}_{ij}^{\;\;\;h}= \tilde{\Gamma}_{ij}^{\;\;\;h} + \frac{1}{2\varphi}\de\varphi/de{x^j}\delta^h_i - \frac{1}{2\varphi}\de\varphi/de{x^p}g^{ph}g_{ij} + \frac{1}{2\varphi}\de\varphi/de{x^i}\delta^h_j
\end{equation}
between the Levi--Civita connections associated respectively to $g_{ij}$ ($\tilde{\Gamma}_{ij}^{\;\;\;h}$) and $\bar{g}_{ij}$ ($\bar{\Gamma}_{ij}^{\;\;\;h}$), it is easily seen that the Klein--Gordon equation expressed in terms of the conformal metric $\bar{g}_{ij}\/$ becomes 
\begin{equation}\label{3.5}
-\de{\psi}/de{x^i}\bar{g}^{ij}\de{\varphi}/de{x^j} + \varphi\bar{\nabla}_j\de{\psi}/de{x^i}\bar{g}^{ij} = m^2\psi
\end{equation}
where $\bar{\nabla}_j\/$ denotes the covariant derivative associated to the conformal metric $\bar{g}_{ij}\/$. Also the trace $\Sigma\/$ can be expressed as function of $\bar{g}_{ij}\/$, that is 
\begin{equation}\label{3.6}
\Sigma=-\de\psi/de{x^p}\de\psi/de{x^q}\varphi\bar{g}^{pq} -2m^2\psi^2\,.
\end{equation}
Now, the relation \eqref{0000.5} links the scalar field $\varphi\/$ to the Klein--Gordon field $\psi\/$, its partial derivatives ${\displaystyle \frac{\partial\psi}{\partial x^i}\/}$ and the conformal metric $\bar{g}_{ij}\/$. 
As already said, in eqs. \eqref{3.1.1} the quantity 
\begin{equation}\label{3.7}
T_{ij} := \frac{1}{\varphi}\Sigma_{ij} -
\frac{1}{\varphi^3}V\/(\varphi)\bar{g}_{ij}
\end{equation}
plays the role of effective stress--energy tensor. The Klein--Gordon equation \eqref{3.2} implies the conservation laws $\tilde{\nabla}^j\Sigma_{ij}=0\/$ and thus also the identities $\bar{\nabla}^jT_{ij}=0\/$ (due to Proposition \ref{Pro3.2}). The latter is a crucial fact which allows to use harmonic coordinates, once again following the same arguments developed in \cite{Wald,yvonne2,yvonne}. 

More in detail, we rewrite the Einstein-like equations \eqref{3.1.1} in the equivalent form
\begin{equation}\label{3.8}
\bar{R}_{ij} = T_{ij} - \frac{1}{2}T\bar{g}_{ij}
\end{equation}
Adopting harmonic coordinates, i.e. local coordinates satisfying the condition
\begin{equation}\label{3.9}
\bar{\nabla}_p\bar{\nabla}^p\/x^i= - \bar{g}^{pq}\bar{\Gamma}_{pq}^i =0\,,
\end{equation}
we can express equations \eqref{3.8} as (see, for example, \cite{Wald,yvonne})
\begin{equation}\label{3.10}
\bar{g}^{pq}\frac{\partial^2 \bar{g}_{ij}}{\partial x^p \partial x^q} = f_{ij}\/(\bar{g},\partial\bar{g},\psi,\partial\psi)
\end{equation}
where $f_{ij}\/$ indicate suitable functions depending only on the metric $\bar{g}\/$, the scalar field $\psi\/$ and their first order derivatives.

Moreover, we assume that  equation \eqref{0000.5} is solvable with respect to the variable $\varphi\/$. In other words, we suppose to be able to derive from eq. \eqref{0000.5} a function of the form
\begin{equation}\label{3.11}
\varphi=\varphi\/(\bar{g},\psi,\de\psi/de{x^p}\de\psi/de{x^q}\bar{g}^{pq})
\end{equation}
expressing the scalar field $\varphi\/$ in terms of the metric $\bar{g}\/$, the Klein--Gordon field $\psi\/$ and its first order derivatives. 
In particular, from eq. \eqref{3.6}, it is easily seen that the dependence of $\varphi\/$ on the derivatives of $\psi\/$ is necessarily of the indicated form. The requirement about  equation \eqref{0000.5} depends on the explicit form of the potential $V\/(\varphi)$. The latter is determined by the function $f(R)\/$ through the formula \eqref{2.8}. Therefore, the above assumption becomes a rule to select viable $f(R)$-models. In addition to this, from eq. \eqref{3.11}, we derive the identity  
\begin{equation}\label{3.12}
\de\varphi/de{x^i}=\de\varphi/de{\left(\de\psi/de{x^s}\de\psi/de{x^t}\bar{g}^{st}\right)}2\de\psi/de{x^q}\bar{g}^{pq}\frac{\partial^2\psi}{\partial x^i \partial x^p} + f_i\/(\bar{g},\partial\bar{g},\psi,\partial\psi)
\end{equation}
Inserting eq. \eqref{3.12} in eq. \eqref{3.5} and taking eqs. \eqref{3.9} into account, we obtain the final form of the Klein-Gordon equation given by 
\begin{equation}\label{3.13}
\left(\bar{g}^{ip} -\frac{2}{\varphi}\de\varphi/de{\left(\de\psi/de{x^s}\de\psi/de{x^t}\bar{g}^{st}\right)}\de\psi/de{x^j}\bar{g}^{ji}\de\psi/de{x^q}\bar{g}^{pq}\right)\frac{\partial^2\psi}{\partial x^i \partial x^p} = f\/(\bar{g},\partial\bar{g},\psi,\partial\psi)
\end{equation}
In eqs. \eqref{3.12} and \eqref{3.13}, $f_i\/$ and $f\/$ indicate suitable functions of $\bar{g}_{ij}\/$, $\psi\/$ and their first order derivatives only. 

Eqs. \eqref{3.10} and \eqref{3.13} describe a second order quasi-diagonal system of partial differential equations for the unknowns $\bar{g}_{ij}\/$ and $\psi\/$. 
The matrix of the principal parts of such a system is diagonal and its elements are the differential operators
\begin{subequations}\label{3.14}
\begin{equation}\label{3.14a}
\bar{g}^{pq}\frac{\partial^2}{\partial x^p \partial x^q}
\end{equation}
and
\begin{equation}\label{3.14b}
\left(\bar{g}^{ip} -\frac{2}{\varphi}\de\varphi/de{\left(\de\psi/de{x^s}\de\psi/de{x^t}\bar{g}^{st}\right)}\de\psi/de{x^j}\bar{g}^{ji}\de\psi/de{x^q}\bar{g}^{pq}\right)\frac{\partial^2}{\partial x^i \partial x^p}
\end{equation}
\end{subequations}
The operator \eqref{3.14a} is noting else but the wave operator associated with the metric $\bar{g}_{ij}\/$, while the operator \eqref{3.14b} is very similar to the sound wave operator arising from the analysis of the Cauchy problem for GR coupled with an irrotational perfect fluid \cite{yvonne4,yvonne}. To discuss the Cauchy problem for system \eqref{3.10} and \eqref{3.13}, we can then follow the same arguments developed in 
\cite{yvonne4,yvonne}.
In particular, if the quadratic form associated with \eqref{3.14b} is of Lorentzian signature and if the characteristic cone of the operator \eqref{3.14b} is exterior to the metric cone, the system \eqref{3.10} and \eqref{3.13} is causal and Leray hyperbolic \cite{Leray,yvonne3}. In such a circumstance, the corresponding Cauchy problem is well-posed in suitable Sobolev spaces. Still following \cite{yvonne4,yvonne}, if the signature of 
$\bar{g}_{ij}\/$ is $(+---)\/$, the required conditions are satisfied when the vector ${\displaystyle \de\psi/de{x^j}\bar{g}^{ij}\/}$ is timelike and 
the inequality
\begin{equation}\label{3.15}
-\frac{2}{\varphi}\de\varphi/de{\left(\de\psi/de{x^s}\de\psi/de{x^t}\bar{g}^{st}\right)}\geq 0
\end{equation}
holds. On the contrary, if the signature of $\bar{g}_{ij}\/$ is $(-+++)\/$, the inequality \eqref{3.15} has to be inverted. As already mentioned, the explicit expression of the function \eqref{3.11} depends on that of the potential \eqref{2.8} which is determined by the function $f(R)\/$. Therefore, the requirement \eqref{3.15} (or, equivalently, its opposite) can be a criterion to single out viable $f(R)\/$-models.

As an example, let us consider again the model $f\/(R)=R+\alpha\/R^2\/$. From the identities $F^{-1}\/(X)=f'\/(X)X -2f\/(X)=-X\/$ and ${\displaystyle (f')^{-1}\/(\varphi)=\frac{\varphi -1}{2\alpha}\/}$ as well as the definition \eqref{2.8}, we get the expression of the effective potential
\begin{equation}\label{4.1.1}
V\/(\varphi)= \frac{1}{8\alpha}(\varphi -1)^2\varphi
\end{equation}
Eq. \eqref{4.1.1} together with eqs. \eqref{0000.5} and \eqref{3.6} yield
\begin{equation}\label{4.2.1}
\varphi = \frac{\left( \frac{1}{2\alpha} + 2m^2\psi^2 \right)}{\left( \frac{1}{2\alpha} - \de\psi/de{x^s}\de\psi/de{x^t}\bar{g}^{st} \right)}
\end{equation}
which expresses the scalar field $\varphi\/$ as function of the metric $\bar{g}_{ij}\/$, the Klein-Gordon field $\psi\/$ and its first order derivatives. Directly from \eqref{4.2.1} it follows
\begin{equation}\label{4.3.1}
\de\varphi/de{\left(\de\psi/de{x^s}\de\psi/de{x^t}\bar{g}^{st}\right)}=  \frac{\varphi}{\left( \frac{1}{2\alpha} - \de\psi/de{x^s}\de\psi/de{x^t}\bar{g}^{st} \right)}
\end{equation}
In the signature $(+---)\/$ for the metric $\bar{g}_{ij}\/$, it is immediately seen that the requirement \eqref{3.15} is automatically satisfied if $\alpha <0\/$ and if ${\displaystyle \bar{g}^{pq}\de\psi/de{x^q}\/}$ is a time-like vector field. 
If the signature is $(-+++)\/$, the condition becomes $\alpha >0\/$. 

\section{Discussion and conclusions}
In this paper, we have discussed   the initial value problem for
 $f(R)$-gravity  in metric and metric-affine approaches.  Both the well-formulation and the well-posedness hold in metric formulation. This result is easily 
 achieved as soon as the theory is recast in term of the O'Hanlon theory. In this case, by choosing a suitable gauge, the Bruhat arguments  for GR can be easily reproduced for any form of  matter acting as source
(e.g. perfect fluid, electromagnetic, Klein-Gordon,Yang-Mills fields). In vacuum case, $f(R)$-gravity is equivalent to GR plus cosmological constant. 
 
On the other hand, it can be shown that the initial value problem
is, in general,  well-formulated also for metric-affine $f(R)$-gravity, at least in presence of the standard matter sources mentioned above \cite{CV3,olmonew}. 
This means that there are no objections to the viability of metric-affine
$f(R)$-gravity, based on the supposed ill-formulation of the Cauchy problem \cite{faraoni}.  However, also the well-posedness is
necessary in order to achieve the complete control of dynamics but, in this case, the role of source fields and the specific form of the function $f(R)$ have to be carefully discussed. 

At present, no general results exist about the well-posedness of the Cauchy problem for metric-affine $f(R)$-gravity. Therefore, to discuss the well-posedness,  we are forced to proceed through  model by model analysis. 
For example, anytime the trace of the stress-energy tensor is constant, the theory reduce to GR plus cosmological constant and the well-posedness is thus recovered, independently of the explicit form of the function $f(R)$. In case of coupling with a perfect-fluid or a Klein-Gordon scalar field, sufficient conditions can be stated in order to assure the well-posedness. It has been proved that there exist some functions $f(R)$ that actually satisfy such conditions which then become a sort of selection rules to determine the  viable forms of $f(R)$-models.  This feature is, in some sense, a sort of consistency check based on a first principle: the initial value problem selects viable models. From a physical viewpoint,  this fact is crucial to discriminate among  competing theories that have to be compared with data.

\end{document}